\documentclass[preprint,pra,superscriptaddress,amsmath,amssymb]{revtex4-1}
\usepackage{amsmath}
\usepackage{amsfonts}
\usepackage{amssymb}
\usepackage{graphicx}
\usepackage{color}
\usepackage{txfonts}
\usepackage[titletoc]{appendix}
\usepackage[colorlinks={true}]{hyperref}
\hypersetup{citecolor={blue}, filecolor={blue}, linkcolor={blue}, urlcolor={blue}}
\begin{document}
\preprint{}
\title{Optimal three-state field-free molecular orientation with terahertz pulses}
\author{Qian-Qian Hong}
\affiliation{Hunan Key Laboratory of Super-Microstructure and Ultrafast Process, School of Physics and Electronics, Central South University,
Changsha 410083, China}
\author{Li-Bao Fan}
\affiliation{Hunan Key Laboratory of Super-Microstructure and Ultrafast Process, School of Physics and Electronics, Central South University,
Changsha 410083, China}
\author{Chuan-Cun Shu}
\email{cc.shu@csu.edu.cn}
\affiliation{Hunan Key Laboratory of Super-Microstructure and Ultrafast Process, School of Physics and Electronics, Central South University,
Changsha 410083, China}
\author{Niels E. Henriksen}
\affiliation{Department of Chemistry, Technical University of Denmark, Building 207, DK-2800 Kgs. Lyngby, Denmark}
 \begin{abstract}
 We present a combined analytical and numerical investigation to show how an optimal control field can be designed to generate maximum field-free orientation of molecules for three populated rotational states. Based on a model involving pure rotational ladder-climbing excitation between rotational states,  a set of optimal amplitude and phase conditions are analytically derived for the applied control fields. The maximum
degree of orientation can be achieved when the field satisfies amplitude and phase conditions at the two transition frequencies. Multiple optimal solutions exist and to examine these conditions, we  devise a quantum coherent control scheme using two terahertz pulses and successfully apply it to the linear polar molecule HCN at ultracold temperature. The sensitivity  of  both populations and phases of rotational states to control field parameters, i.e., the detuning, bandwidth, and time delay,  is analyzed for understanding the optimal orientation mechanism. This work thus examines the frequency domain landscape belonging to optimal pulses. 
\end{abstract}
\maketitle
\section{Introduction}
Controlling molecular rotations associated with molecular alignment and orientation is an active research area of experimental and theoretical molecular science\cite{qr-TS,IRPC2010,IJC2012,qr-sugny,NP-20}, because of its fundamental importance in physics and chemistry. An oriented molecular sample has an asymmetric angular distribution upon reflection, while this distribution for the aligned molecule is symmetric but not isotropic. As a result,  achieving molecular orientation since it requires the breakdown of inversion symmetry is more challenging than the well-established alignment in both the adiabatic and non-adiabatic regimes, where  the adiabatic limit implies that the duration of the applied laser pulses is much longer than the characteristic timescale of the free
rotation of the molecule. Over the years, extensive efforts in theory and experiment have been put into generating a post-pulse orientation - known as the field-free orientation of molecules \cite{qr-sakai,qr-mjj,qr-kling,PRL:2014,qr-wj,PRA-208,PRA-2020}. It has potential applications ranging from molecular-phase modulators, ultrafast X-ray diffraction, and ultrashort pulse compression to chemical reactivity, nanoscale design, high harmonic generation, and molecular rotational echoes \cite{jacs2009,wj2,science2013,nm,HHG-LU,mcm1,PRL-2015,PRL-2019}. \\ \indent  
The physical mechanism for generating the field-free orientation is to create a rotational wave packet, which consists of a superposition of rotational states with even and odd angular momentum quantum numbers. Experimentally, all-optical techniques in generating such a  field-free orientation have become mature thanks to the non-resonant light-interactions with molecular polarizability, and hyperpolarizability \cite{qr-sakai,qr-mjj,qr-kling,PRL:2014,qr-wj}. As compared with the all-optical techniques, the progress in the realization of the field-free orientation of molecules exclusively using resonant terahertz pulses was relatively slow \cite{ex0,ex00} since it was proposed 20 years ago \cite{niels3,qr-niels2,dion}. The technical difficulty was previously attributed to using a highly asymmetric half-cycle terahertz pulse, which impulsively transfers angular momentum to the molecule due to a non-zero (time-integrated) area of the  short central part \cite{Sugny4,sugny2,pra2006,shu2,shu3,shuJCP}. Further theoretical investigations have shown that molecules can be oriented through resonant excitation using symmetric terahertz pulses with a zero-time-integrated area. In addition, applying an intense non-resonant ultrashort pulse to align the molecule prior to the resonant terahertz excitation, comprehensive theoretical research has recently resulted in the experimental demonstration of enhanced field-free orientation in the sudden-impact limit \cite{jiro,shu4,ex1,ex2}. However, it remains challenging to obtain a robust field-free orientation without the additional use of an intense non-resonant pulse or a static electric field.\\ \indent
Despite the slow progress in experiments, several theoretical proposals have been directed toward the realization of molecular field-free orientation by exclusively using zero-area terahertz pulses \cite{J2,Sugny6,Sugny2014,JCP2017,Sugny2019}. Furthermore, recent theoretical work has shown that a single half-cycle-zero-area terahertz pulse can result in steady molecular orientation long time after the pulse is turned off \cite{PRL2020}.  We recently examined a three-state molecular orientation model with an experimentally available single-cycle terahertz pulse \cite{shu2020}. We found that the theoretical maximum degree of orientation for three populated rotational states can be obtained in the intermediate nonadiabatic limit. The corresponding rotational excitation processes consist of rotational ladder-climbing and simultaneous multiphoton absorption. It is a fundamentally interesting question whether the pure rotational ladder-climbing excitation can independently generate the maximum orientation while suppressing simultaneous multiphoton absorption processes.  \\ \indent
In this work, we perform a further investigation of the three-state model. Based on an analytical wave function for a three-state system, we derive amplitude and phase conditions for the control fields, capable of generating the optimal degree of orientation for three populated rotational states. We then devise a quantum coherent control scheme to examine our theoretical analysis using two terahertz pulses and apply it to the linear polar molecule HCN with four representative simulations. We find that the pure rotational ladder-climbing excitation can induce the maximum degree of orientation as long as the pulses satisfy amplitude and phase conditions at the two transition frequencies. This work provides a deep insight into optimal molecular rotations by involving the rotational ladder-climbing excitation mechanism. It also offers an essential reference for designing an experimental scheme toward realizing optimal field-free orientation within a finite rotational Hilbert subspace. 
\\ \indent
The remainder of this paper is organized as follows.  In Sec. \ref{methods}, we describe the theoretical methods for analyzing the three-state-orientation model. We present the results of the numerical simulations and discussion in Sec. \ref{RandD}. Finally, we conclude with a brief summary in Sec. \ref{conl}.
\section{ Theoretical Methods} \label{methods}
We consider a general model for producing three populated rotational states  by using  linearly polarized terahertz pulses $\mathcal{E}(t)$,  which turns on at $t_0$ and off at $t_f$ with $\mathcal{E}(t_0)=\mathcal{E}(t_f)=0$. For the molecule in its electronic and vibrational ground state,  the molecular Hamiltonian can be given by $\hat{H}(t)=\hat{H_0}+\hat{V}(t)$, where $\hat{H}_0=B\hat{J}^2$ is the field-free Hamiltonian for a linear molecule with the angular momentum operator $\hat{J}$, rotational constant $B$ and $\hat{V}(t)$ denotes the interaction Hamiltonian. Within the dipole approximation, the interaction Hamiltonian can be given by  $\hat{V}(t)=-\mu \mathcal{E}(t)\cos\theta$, where $\theta$ is the angle between the rotor axis and the pulse polarization. Note that we consider the electric-field strengths 
of $\mathcal{E}(t)$ in a relatively weak regime. The contribution of the molecular polarizability and hyperpolarizability to the rotational excitation can be ignored.\\ \indent  The time evolution of the system in the interaction picture from the initial time $t_0$ to a given time $t$ can be described by a unitary operator  $\hat{U}(t,t_0)$, which has a solution ($\hbar=1$)
\begin{eqnarray} 
\hat{U}(t, t_0)=\mathbb{I}-i\int_{t_0}^t dt'\hat{H}_I(t')\hat{U}(t',t_0)\label{UO}
\end{eqnarray}
where  $\hat{H}_I(t)=\exp(i\hat{H}_0t)[-\hat{\mu}\mathcal{E}(t)]\exp(-i\hat{H}_0t)$, and the matrix elements of the dipole operator $\hat{\mu}$ reads $\mu_{JJ'}=\mu\langle J'M|\cos\theta|JM\rangle$ with $\langle J+1M|\cos\theta|JM\rangle=\sqrt{(J+1)^2-M^2}/\sqrt{(2J+1)(2J+3)}$.  The time-dependent three-state wave packet after the laser pulse excitation is given by 
 \begin{eqnarray}\label{wf}
|\psi_{J_{0}M}(t)\rangle=\sum_{J'=0}^2c_{J'}(t_f)e^{-iE_{J'M}t}|J'M\rangle
\end{eqnarray}
where the rotational eigenstates $|J'\rangle$ satisfy $\hat{H_{0}}|J'\rangle=E_{J'}|J'\rangle$ with eigenenergies  $E_{J'}=BJ'(J'+1)$, and $c_{J'}$ are  the expansion coefficients of $|J'\rangle$, which  can be obtained by $c_{J'}(t)=\langle J'M|\hat{U}(t,t_0)|J_{0}M\rangle$ with the initial state $|J_0M\rangle$. The degree of orientation after the rotational excitation with the selection rule $\Delta J=\pm1$ can be written as
\begin{eqnarray} \label{cos}
\left\langle\cos\theta\right\rangle(t)&=&2\mathcal{M}_{1,0}\left|c_{1}\left(t\right)\right|\left|c_{0}\left(t\right)\right|\cos\left(\omega_{01}t-\phi_{01}\right)\\ \nonumber
&&+2\mathcal{M}_{2,1}\left|c_{2}\left(t\right)\right|\left|c_{1}\left(t\right)\right|\cos\left(\omega_{12}t-\phi_{12}\right)
\end{eqnarray}
where $\mathcal{M}_{J',J}=\langle J'M|\cos\theta|JM\rangle$, the transition frequencies are defined by $\omega_{01}=(E_{1}-E_{0})=2B$ and $\omega_{12}=(E_{2}-E_{1})=2\omega_{01}=4B$,  and the relative phases are $\phi_{01}=\arg[c_{1}(t)]-\arg[c_{0}(t)]$ and $\phi_{12}=\arg[c_{2}(t)]-\arg[c_{1}(t)]$. It shows that full revivals in the field-free orientation occurs at a time interval $\tau=\pi/B$ when $\phi_{12}=2\phi_{01}+k\pi$ $(k=0, \pm1, \pm2, \dots)$ by
generating the coherent superposition of rotational states.\\ \indent
\subsection{The maximum degree of orientation for a three-state model}
 Based on the method of Lagrange multiplies, the maximum degree of orientation with the three state subspace  can be estimated by  \cite{Sugny:PRA:2004,NEH2020} 
\begin{eqnarray} \label{LE}
\mathcal{L}(|c_0|, |c_1|, |c_2|, \lambda)=f-\lambda g
\end{eqnarray}
where $f=2\mathcal{M}_{1,0}|c_1c_0|+2\mathcal{M}_{2,1}|c_2c_1|$ corresponds to the amplitude of the orientation at the full revivals by considering the relative phases  in Eq. (\ref{cos}) to satisfy a relation of  $\phi_{12}=2\phi_{01}+k\pi$ $(k=0, \pm1, \pm2, \dots)$, and $g=|c_0|^2+|c_1|^2+|c_2|^2-1=0$ is a constraint. The extremum of $f$ subject to $g$ can be obtained by satisfying
 $\bigtriangledown\mathcal{L}=0$, then we have 
 \begin{eqnarray} \label{LE1}
\mathcal{M}_{1,0}|c_1|-\lambda|c_0|&=&0 \\  \nonumber
\mathcal{M}_{2,1}|c_2|+\mathcal{M}_{1,0}|c_0|-\lambda|c_1|&=&0 \\  \nonumber
\mathcal{M}_{2,1}|c_1|-\lambda|c_2|&=&0.
\end{eqnarray}
By multiplying $|c_0|$, $|c_1|$ and $|c_2|$ to each sub-equation in Eq. (\ref{LE1}), respectively, we have 
 \begin{eqnarray} \label{LE2}
f-\lambda(|c_0|^2+|c_1|^2+|c_2|^2)=f-\lambda=0.
\end{eqnarray}
The maximum degree of orientation $f$ corresponds to  the maximum value of $\lambda$ governed by Eq. (\ref{LE1}).  By multiplying $|c_0|$, $-|c_1|$, and $|c_2|$  for the first,  second, and third sub-equation in Eq. (\ref{LE1}), respectively, we can obtain a relation of $|c_0|^2+|c_2|^2=|c_1|^2$ when the maximum degree of orientation $f$ is reached. For the molecules initially in the ground rotational state with $J=0$ and $M=0$, we can calculate the matrix elements of $\mathcal{M}_{1,0}=\sqrt{1/3}$ and $\mathcal{M}_{2,1}=\sqrt{4/15}$. As a result, the maximum degree of orientation can be obtained by 
\begin{eqnarray} \label{MOD}
\lambda=\frac{\mathcal{M}_{2,1}|c_2|+\mathcal{M}_{1,0}|c_0|}{|c_1|}
\end{eqnarray}
which reaches its maximum 0.7746 with $|c_0|=\sqrt{10}/6$, $|c_1|=\sqrt{2}/2$ and $|c_2|=\sqrt{2}/3$, corresponding to three populated rotational states with $p_0=|c_0|^2=0.278$, $p_1=|c_1|^2=0.5$ and $p_2=|c_2|^2=0.222$. 
\subsection{An analytical solution for the rotational wave packet}
We now present a theoretical  analysis to show the dependence of $c_{J'}(t)$ on the control field $\mathcal{E}(t)$. We write the Hamiltonian in the interaction picture without using the rotating wave approximation \cite{pra:92:063815,shupra,shuprl}
\begin{equation} \label{HI}
\hat{H}_{I}(t)=-\left(\begin{array}{ccc}
0 & \mu_{10}\mathcal{E}(t)e^{-i\omega_{01}t} & 0\\
\mu_{10}\mathcal{E}(t)e^{i\omega_{01}t} & 0 & \mu_{21}\mathcal{E}(t)e^{-i\omega_{12}t}\\
0 & \mu_{21}\mathcal{E}(t)e^{i\omega_{12}t} & 0
\end{array}\right). 
\end{equation}
We expand the unitary operator $\hat{U}(t, t_0)$ by using the Magnus expansion \cite{pr:470:151}
\begin{equation}
\hat{U}(t,t_0)=\exp\Bigg[\sum_{n=1}^{\infty}\hat{S}^{(n)}(t)\Bigg]
\end{equation}
where the first  leading term is given by 
\begin{equation}
\hat{S}^{(1)}(t)=-i\int_{t_0}^{t}dt_1\hat{H_I}(t_1).
\end{equation}
The corresponding time-dependent wave function $|\psi^{(1)}(t)\rangle\equiv\sum_{J'=0}^2c_{J'}^{(1)}|J'0\rangle=\hat{U}^{(1)}(t, t_0)|00\rangle$  starting from the ground rotational state $|00\rangle$ can be given by \cite{shu2020}
 \begin{eqnarray} \label{J0M0}
|\psi^{(1)}(t)\rangle&=&\frac{[\left|\theta_2(t)\right|^2+\left|\theta_1(t)\right|^2\cos\theta_{12}(t)]}{\theta_{12}^2(t)}|00\big\rangle \\ \ \nonumber
&&+\frac{i\theta_1(t)\sin\theta_{12}(t)}{\theta_{12}(t)}|10\rangle+\frac{\theta_1(t)\theta_2(t)}{\theta_{12}^2(t)}\left[\cos\theta_{12}(t)-1\right]|20\rangle,
\end{eqnarray}
where $\theta_{12}(t)=\sqrt{|\theta_{1}(t)|^{2}+|\theta_{2}(t)|^{2}}$ with 
\begin{eqnarray} \label{theta1}
\theta_{1}(t)=\mu_{10}\int_{t_{0}}^{t}dt'\mathcal{E}(t')e^{i\omega_{01}t'}
\end{eqnarray}
and 
\begin{eqnarray} \label{theta2}
\theta_{2}(t)=\mu_{21}\int_{t_{0}}^{t}dt'\mathcal{E}(t')e^{i\omega_{12}t'}. 
\end{eqnarray}\\ \indent
 Thus, $\theta_{i}(t)$ is the transition dipole moment times the Fourier transform of the electric field at the transition frequency. A rotational ladder-climbing mechanism is involved in generating the superposition of rotational states in Eq. (\ref{J0M0}) \cite{JCP2005}. That is, the molecule is excited resonantly  by a one-photon transition from $|00\rangle$ to $|10\rangle$ followed by a one-photon transition  from $|10\rangle$ to $|20\rangle$. 
\subsection{The amplitude and phase conditions}
 Based on the above analysis,  the maximum degree of orientation in Eq. (\ref{MOD}) requires that $\theta_1(t_f)$ and $\theta_2(t_f)$ satisfy the following relations
 \begin{eqnarray} \label{amp}
\Big|c_0^{(1)}(t_f)\Big|&\equiv&\Bigg|\frac{\left|\theta_2(t_f)\right|^2+\left|\theta_1(t_f)\right|^2\cos\theta_{12}(t_f)}{\theta_{12}^2(t_f)}\Bigg|=\frac{\sqrt{10}}{6}\\ \nonumber
\Big|c_1^{(1)}(t_f)\Big|&\equiv&\Bigg|\frac{i\theta_1(t_f)\sin\theta_{12}(t_f)}{\theta_{12}(t_f)}\Bigg|=\frac{\sqrt{2}}{2}\\ \nonumber
\Big|c_2^{(1)}(t_f)\Big|&\equiv&\Bigg|\frac{\theta_1(t_f)\theta_2(t_f)}{\theta_{12}^2(t_f)}\left[\cos\theta_{12}(t_f)-1\right]\Bigg|=\frac{\sqrt{2}}{3}.
\end{eqnarray}
From Eq. (\ref{amp}), we can derive  
\begin{eqnarray} 
\sum_{\ell=0}^4a_\ell\Bigg[\frac{|\theta_2(t_f)|}{|\theta_1(t_f)|}\Bigg]^{\ell}=0\label{xx}
\end{eqnarray}
with the coefficients $a_0=2/9$, $a_1=a_3=-2\sqrt{2}/3$, $a_2=17/18$, and $a_4=13/18$.\\ \indent 
By setting $s=|\theta_2(t_f)|/|\theta_1(t_f)|$, there exists two real solutions to Eq. (\ref{xx}) with $s_1=0.9967$ and $s_2=0.3087$. To that end, we obtain two conditions for generating the maximum orientation, i.e., 
\begin{eqnarray} \label{ac1}
|\theta _{1} ( t_{f}  )| =\frac{\big | \arccos( 1-\frac{\sqrt{2} }{3s_{1}}-\frac{\sqrt{2}s_{1}}{3})+2j\pi \big| }{\sqrt{1+s_{1}^{2}  } } , \ \  |\theta_2(t_f)|=s_{1}|\theta_{1}(t_f)|, 
\end{eqnarray} and 
\begin{eqnarray} \label{ac2}
|\theta_1(t_f)|=\frac{\big | \arccos( 1-\frac{\sqrt{2} }{3s_{2}}-\frac{\sqrt{2}s_{2}}{3})+2j\pi \big| }{\sqrt{1+s_{2}^{2}  } }, \ \  |\theta_2{t_f}|=s_{2}|\theta_{1}(t_f)|, 
\end{eqnarray}
 where $j=0, \pm1, \pm2, \cdots$. To make the relative phases in Eq. (\ref{cos}) to meet the relation of  $\phi_{12}=2\phi_{01}+k\pi$ $(k=0, \pm1, \pm2, \dots)$, 
  $\theta_1(t_f)$ and $\theta_2(t_f)$ in Eq. (\ref{J0M0}) are required to  satisfy the  phase condition
\begin{eqnarray} \label{phase}
2\arg[\theta_{1}(t_f)]-\arg[\theta_{2}(t_f)]=\Bigg(k\pm\frac{1}{2}\Bigg)\pi, \ \  
\end{eqnarray}\\ \indent
 Note that the theoretical maximum value for the degree of orientation is independent of the excitation scheme. The above amplitude and phase conditions only work to generate the rotational wave packet subject to the first-order Magnus description in Eq. (\ref{J0M0}).
\subsection{A coherent control scheme with the use of two terahertz pulses}
To use the above amplitude and phase conditions, we design a quantum coherent control scheme by using two terahertz pulses. From the frequency domain of view, the complex values of $\theta_1(t_f)$ and $\theta_2(t_f)$ are related to the Fourier transform of the control field $\mathcal{E}(t)$, \begin{eqnarray} \label{FT}
E(\omega)\equiv A(\omega)e^{i\phi(\omega)}=\int_{t_0}^{t_f}dt'\mathcal{E}(t')e^{-i\omega t'}.
\end{eqnarray}
 where $A(\omega)$ and $\phi(\omega)$ denotes the spectral amplitude and spectral phase of the field, respectively. By comparing Eq. (\ref{FT}) with Eqs. (\ref{theta1}) and (\ref{theta2}), we can obtain two  relations
 \begin{eqnarray} \label{theta11}
\theta_{1}^{*}(t_f)=\mu_{10}A(\omega_{01})e^{i\phi(\omega_{01})}
\end{eqnarray}
and 
\begin{eqnarray} \label{theta22}
\theta_{2}^{*}(t_f)=\mu_{21}A(\omega_{12})e^{i\phi(\omega_{12})}.
\end{eqnarray}
 Thus, the amplitude and phase conditions can be satisfied by optimizing the values of $A(\omega_{01})$, $A(\omega_{12})$,  $\phi(\omega_{01})$, $\phi(\omega_{12})$ and all pulses that satisfies Eqs. (\ref{ac1})-(\ref{phase}) will give the maximum degree of orientation within the three-state model. \\ \indent 
\begin{figure}[!t]\centering
\resizebox{0.6\textwidth}{!}{%
 \includegraphics{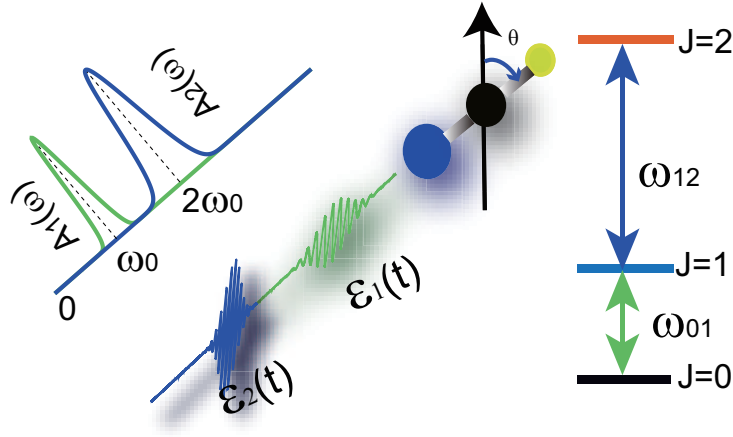}
} \caption{Schematic of rotational excitation of a linear polar molecule within a three-state model by using two terahertz pulses. The first pulse $\mathcal{E}_1(t)$ centered at $t=0$ with a fundamental frequency $\omega_0$ and the second pulse with a time delay $\tau_0$ and a center frequency of $2\omega_0$. $A_1(\omega)$ and $A_2(\omega)$ correspond to the spectral amplitudes of the pulses. $\theta$ denotes the angle between the rotor axis and the polarization of the pulsed fields. The three-state model consists of states with rotational quantum numbers of $J=0$, 1 and 2. $\omega_{01}$ and $\omega_{12}$ are the transition frequencies between states. } \label{fig1}
\end{figure}
To be specific, we use two Gaussian terahertz pulses to design the complex spectral field  by
\begin{eqnarray} \label{Aw}
E(\omega)=\sum_{i=1,2}A_ie^{-\frac{(\omega-\omega_i)^2}{2\Delta\omega_i^2}}e^{i\phi_i(\omega)}e^{-i\omega\tau_i},
\end{eqnarray}
where $A_i$, $\omega_i$, $\Delta\omega_i$, $\phi_i$ and $\tau_i$ denotes the amplitude, fundamental frequency, bandwidth, spectral phase and center time of $i$th pulse, respectively. By setting the values of $\omega_1$ and $\omega_2$ to the transition frequencies $\omega_{01}$ and $\omega_{12}$, the optimal amplitudes of $\theta_1(t_f)$ and $\theta_2(t_f)$ can be obtained by scaling  the amplitudes $A_1$ and $A_2$, and the phase conditions can be satisfied by modulating the combined values of $\phi_1(\omega_{01})-\omega_{01}\tau_1$ and $\phi_2(\omega_{12})-\omega_{12}\tau_2$.  Note that the maximum orientation is only obtained when the two pulses do not overlap in frequency space, and therefore two narrow-bandwidth terahertz pulses consisting of multiple cycles will be required to satisfy the criteria of  Eqs. (\ref{ac1}) and (\ref{ac2}). It implies that the present two-pulse scheme is different from our previous work \cite{shu2020} using a single-cycle terahertz pulse, for which the rotational excitation involves optical processes governed by high-order Magnus terms, going beyond the present rotational ladder-climbing mechanism.  
 \section{Results and discussion} \label{RandD}
 We now perform simulations to examine the above model by using two terahertz pulses,  as illustrated in Fig. \ref{fig1},  which is applied to the linear polar molecule HCN. The used pulses are comprised of the fundamental frequency $\omega_1=\omega_0$ and the second harmonic frequency $\omega_2=2\omega_0$. We consider two pulses with the same bandwidth $\Delta\omega_1=\Delta\omega_2=\Delta\omega$.  The optimal amplitudes of $A_1$ and $A_2$  are chosen with  $A_1=|\theta_1(t_f)|/\mu_{10}$ and $A_2=|\theta_2(t_f)|/\mu_{21}$.  We fix the center time of the first pulse at $\tau_1=0$ and set the second one $\tau_2=\tau_0$. Thus, the time-dependent electric field $\mathcal{E}(t)$ can be given by 
\\ \indent
\begin{eqnarray} \label{tde1}
\mathcal{E}(t)=\frac{1}{\pi}\mathcal{R}\Bigg\{\int_{0}^{\infty}d\omega\Bigg[\frac{|\theta_1(t_f)|}{\mu_{10}}e^{-\frac{(\omega-\omega_0)^{2}}{2\Delta\omega^2}}e^{i\phi_1}
+\frac{|\theta_2(t_f)|}{\mu_{21}}e^{-\frac{(\omega-2\omega_0)^2}{2\Delta\omega^2}}e^{i\phi_2}e^{-i\omega\tau_0}\Bigg]e^{i\omega t} \Bigg\}.
\end{eqnarray}\\ \indent
Based on the above analysis, we find that the spectral amplitude and phase at the two transition frequencies is all that matters. The frequency-domain shaping analysis from Eq. (\ref{FT}) to Eq. (\ref{tde1}) helps us understand how to design the control fields to satisfy the amplitude and phase conditions. To that end, we can give  the time-dependent control field in Eq. (\ref{tde1}) by 
\begin{eqnarray}
\mathcal{E}(t)=\sqrt{\frac{2}{\pi}}\frac{1}{\tau}\Bigg\{\frac{|\theta _{1}(t_f) |}{\mu _{10}}e^{-\frac{t^{2} }{2\tau^2} }\cos\left ( \omega _{0}t +\phi _{1} \right )+\frac{|\theta _{2}(t_f)| }{\mu _{21}} 
e^{-\frac{(t-\tau_0)^{2} }{2\tau^2} }\cos\left [ 2\omega_0(t-\tau_0) +\phi _{2}   \right ]\Bigg\},
\end{eqnarray}\\
where the pulse duration is defined by $\tau=1/\Delta\omega$.  Thus the control scheme shown in Fig. \ref{fig1} corresponds to the commonly used pump-pump control in experiments involving two time-delayed pulses comprised of the fundamental frequency and the second harmonic frequency. The amplitude and phase conditions in Eqs. (\ref{ac1})-(\ref{phase}) can be fulfilled by controlling the values of the amplitude and center frequency of the pulses while fixing the phase. That is,  the center frequency, strength, absolute phase of the laser fields can be used as the control parameters to optimize the maximum degree of orientation.
\begin{figure}[!t]\centering
\resizebox{0.7\textwidth}{!}{%
 \includegraphics{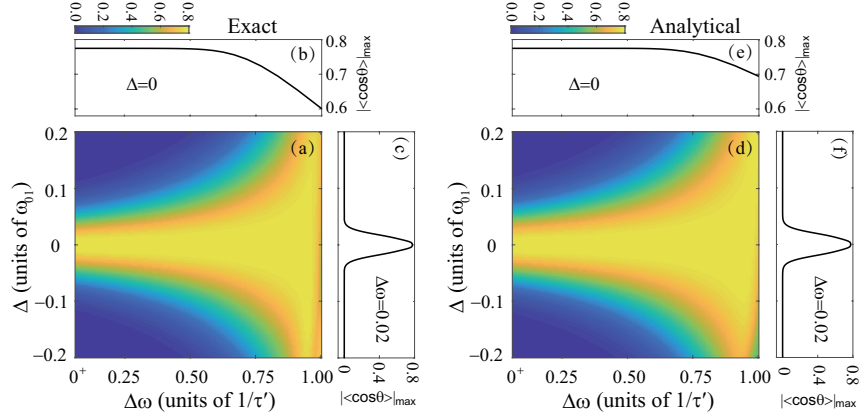}
} \caption{Numerical simulations with two zero-delayed pulses using the condition in Eq. (\ref{ac1}). (a) and (b) The maximum orientation values $\left|\langle\cos\theta\rangle\right|_{max}$ versus the bandwidth and the detuning of the pulses. The left panels (a-c) correspond to the exact simulations by using the time-dependent unitary operator defined by Eq. (\ref{UO}). The right panels (d-f) correspond to the analytical simulations by using Eq. (\ref{J0M0}). $1/\tau'=2\omega_{01}/\pi$ used as the unit corresponds to the bandwidth of a single-cycle pulse, i.e., the shortest possible pulse of a given wavelength. }\label{fig2}
\end{figure} 
\subsection{Simulations with two zero-delayed pulses}
We first examine the case of the time delay $\tau_0=0$, i.e., the two pulses act on the molecule simultaneously. The first amplitude condition  of Eq. (\ref{ac1}) is used by considering $j=0$, i.e., $|\theta_1(t_f)|=0.3412\pi$ and $|\theta_2(t_f)|=0.3401\pi$. The phases  are fixed at  $\phi_1=\phi_2=-\pi/2$ by Eq. (\ref{phase}).  Figure \ref{fig2}  shows the dependence of the maximal orientation on the bandwidth  $\Delta \omega$ and the detuning $\Delta=\omega_{0}-\omega_{01}$, in which the exactly calculated results by  Eq. (\ref{UO}) are compared with the analytical results by using the first-order Magnus approximation. Both simulations show the maximum value of the orientation and its dependence on the parameters, and the exact result and the analytical match better and better as the bandwidth becomes narrower. To further show this change, Figs. \ref{fig2} (b) and (e) shows the maximum degree of orientation versus the bandwidth by fixing $\Delta=0$, and Figs. \ref{fig2} (c) and (f) exhibits its dependence on detuning by fixing the bandwidth at $\Delta\omega=0.02/\tau'$ with $\tau'=\pi/2\omega_{01}$.
In the broad bandwidth regime, the maximum orientation value is smaller than the theoretical maximum $0.7746$, and there are some slight differences between the exact and analytical results.  The former can be attributed to the overlap between the frequency distributions of the two pulses. That is, the fundamental-frequency pulse with a broad bandwidths also results in transitions between the states $|10\big\rangle$ and $|20\big\rangle$, and the second harmonic one for the same reason also induces transitions between the states $|00\big\rangle$ and $|10\big\rangle$. The latter is caused by optical transitions beyond the rotational ladder climbing, as discussed in our previous work \cite{shu2020}. The simultaneous two-photon transitions from $|00\big\rangle$ to $|20\big\rangle$ play a role. \\ \indent
\begin{figure}[!t]\centering
\resizebox{0.7\textwidth}{!}{%
 \includegraphics{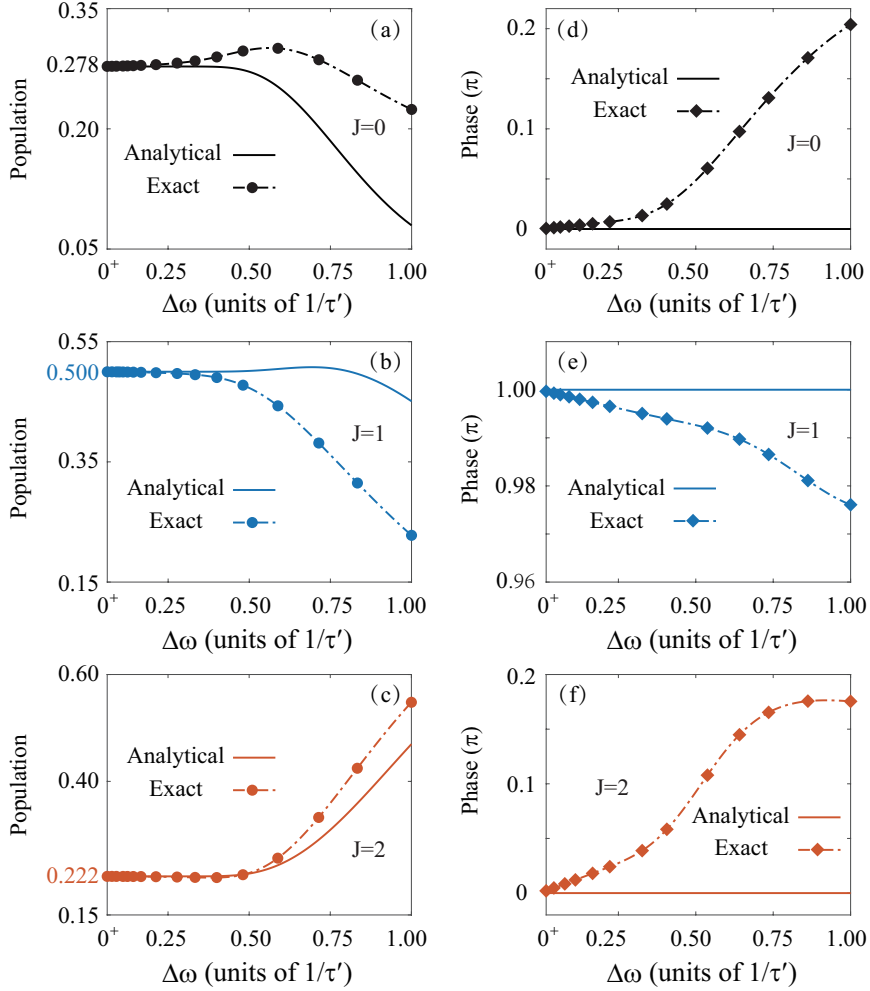}
} \caption{The final populations and phases of the three rotational states $|00\rangle$, $|10\rangle$ and $|20\rangle$ versus the bandwidths of the pulses at a detuning $\Delta=0$. The exact (dash-dotted lines) populations and phases, i.e., $|c_{J'}(t_f)|^2$ and $\arg[c_{J'}(t_f)]$ are compared with the analytical (solid lines) ones , i.e., $|c_{J'}^{(1)}(t_f)|^2$ and $\arg[c_{J'}^{(1)}(t_f)]$.} \label{fig3}
\end{figure}
To gain an insight into the underlying excitation mechanism, Fig. \ref{fig3} shows the final populations and phases versus the bandwidth by setting the detuning $\Delta=0$, which correspond to the orientation in Figs. \ref{fig2} (b) and (e).  The populations and phases of the rotational states are strongly dependent on the pulse bandwidths and gradually converge to the maximum theoretical values by decreasing the pulse bandwidths, leading to the theoretical maximum degree of orientation. We can see that there are visible differences between the exact results and analytical results by using the first-order approximation, indicating that the optical transitions via the higher-order terms in the Magnus expansion occur in the broad-bandwidth regime. However, when the laser pulses turn into the narrow-bandwidth regime, the exact results are in good agreement with the analytical model, indicating that the optical processes via high-order Magnus terms are suppressed.  As a result, the optical transitions via the rotational ladder-climbing described by Eq. (\ref{J0M0}) determine the orientation dynamics in the narrow bandwidth regime.    \\ \indent 
\begin{figure}[!t]\centering
\resizebox{0.7\textwidth}{!}{%
 \includegraphics{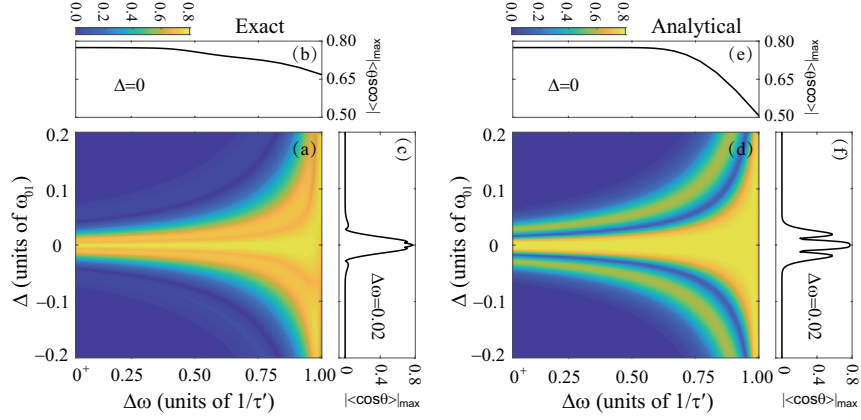}
} \caption{The same simulations as the Fig. \ref{fig2}  for the second condition in Eq. (\ref{ac2}). }\label{fig4}
\end{figure}
\begin{figure}[!t]\centering
\resizebox{0.7\textwidth}{!}{%
 \includegraphics{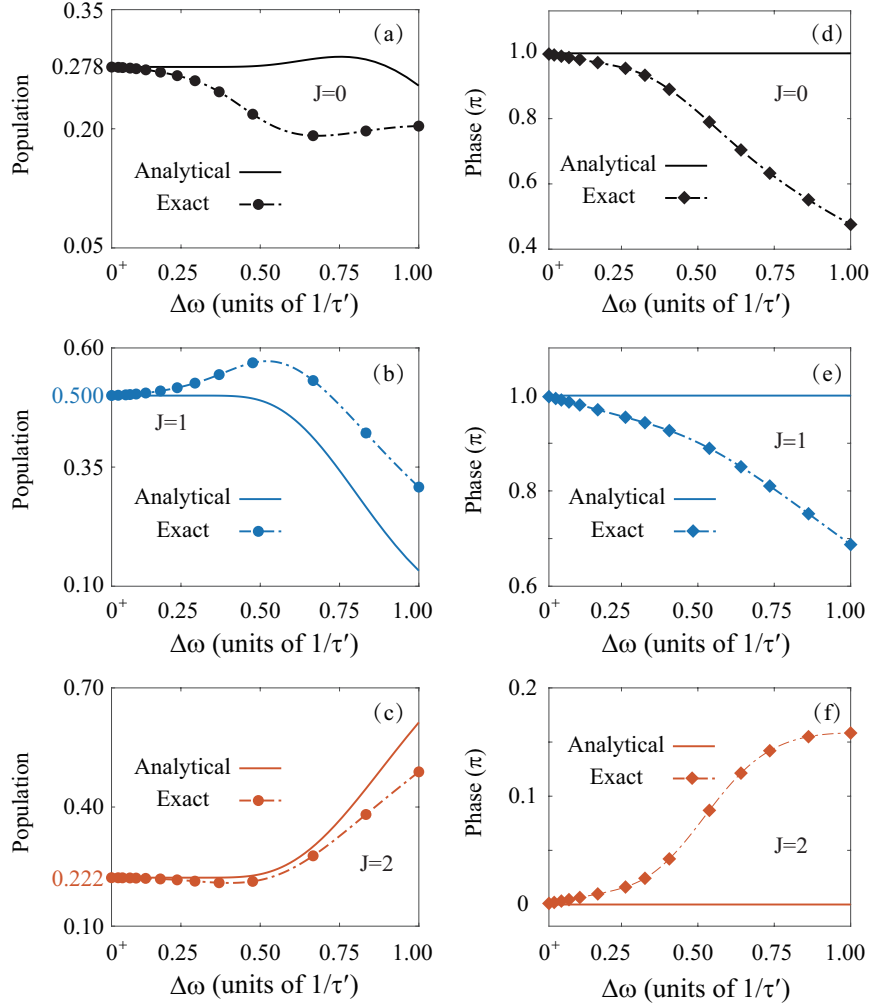}
} \caption{The same simulations as the Fig. \ref{fig3}  for the second condition in Eq. (\ref{ac2})} \label{fig5}
\end{figure}
Figures \ref{fig4} and \ref{fig5} show the same simulations as in Fig. \ref{fig2} and \ref{fig3}, now for the second optimal condition of Eq. (\ref{ac2}) with $j=0$, i.e., $|\theta_1(t_f)|=0.7021\pi$ and $|\theta_2(t_f)|=0.2167\pi$. The results in Figs. \ref{fig4} and \ref{fig5} show similar behaviors like that in Figs. \ref{fig2} and \ref{fig3}, but there is a visible difference between the exact simulations and the first-order analytical ones. The maximum orientation for both simulations is highly sensitive to both detuning and bandwidth. We first examine the case of $\Delta=0$, for which the exact results can reach the theoretical maximum as the analytical one in the narrow bandwidth regime, see details in Figs. \ref{fig4} (b) and (e) and Fig. \ref{fig5}. By comparing the phases of the second condition in Figs. \ref{fig5} (d)-(f) with the first one in Fig. \ref{fig3} (d)-(f), it is interesting to see that the phase of the ground rotational state flips by $\pi$ after the pulse radiations. This phenomenon  can be explained by using  Eq. (\ref{J0M0}) with $\theta_1(t_f)=0.7021\pi$ and $\theta_2(t_f)=0.2167\pi$, which does lead to a phase flip for the ground state $|00\rangle$,  showing an essential difference from that satisfying the first amplitude condition. In the broad bandwidth regime, the differences in both populations and phases become obvious because of the pulse overlap in frequency domain and the optical transitions via high-order terms in the Magnus expansion, which reduces the orientation values below the theoretical maximum. \\ \indent
\begin{figure}[!t]\centering
\resizebox{0.6\textwidth}{!}{%
 \includegraphics{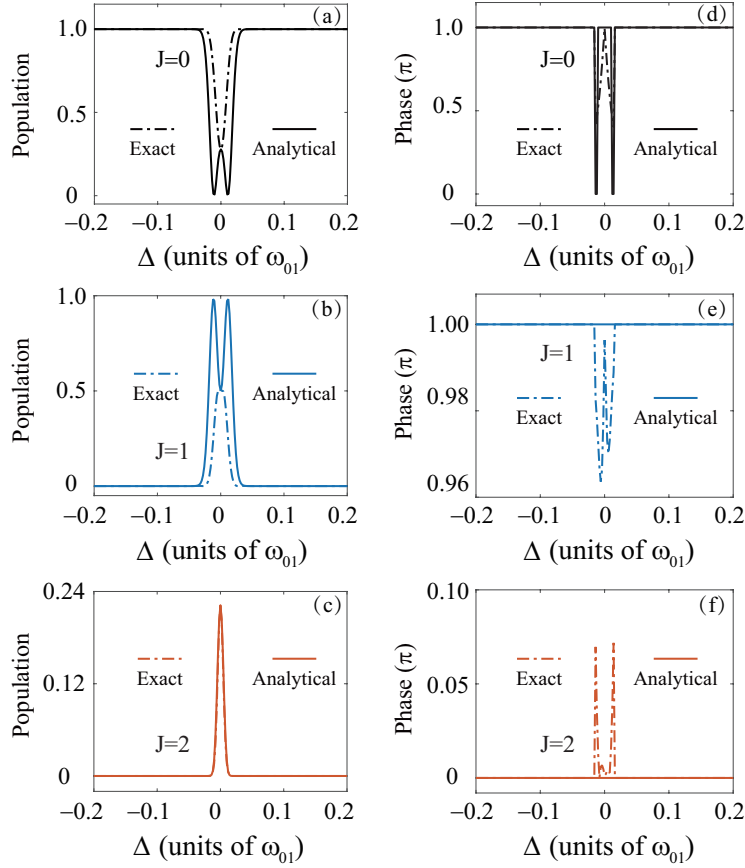}
} \caption{The dependence of populations and phases on the detuning for the control field with a narrow bandwidth of $\Delta\omega=0.02/\tau'$. The dashed lines show the exactly calculated  populations and phases,i.e., $|c_{J'}(t_f)|^2$ and $\arg[c_{J'}(t_f)]$,  and the solid lines correspond to analytical ones, $|c_{J'}^{(1)}(t_f)|^2$ and $\arg[c_{J'}^{(1)}(t_f)]$. Note that the  phases for larger detunings where the populations of the  states $|10\rangle$ and $|20\rangle$ are nearly zero are meaningless and therefore are set to zero in our simulations.} \label{fig6}
\end{figure}
As shown in Fig. \ref{fig4}, the analytically calculated value of the orientation shows strong oscillations as a function of detuning even in the narrow bandwidth regime. However, this phenomenon is not clearly visible in the exact simulations. Figure \ref{fig6} shows the population and phase dependence on the detuning at $\Delta\omega=0.02/\tau'$, corresponding to the orientations in Figs. \ref{fig4} (c) and (f). We can see that the exactly calculated population against detuning changes for the state $|20\rangle$ follows the analytical one well. The visible difference in phases implies that the optical transition processes via high-order Magnus terms play a role even in the narrow bandwidth region,  as shown in Fig. \ref{fig6} (f).  For the state $|10\rangle$, the analytically calculated population shows strong Rabi oscillations concerning the detuning, which is caused by a larger Rabi coupling between the states $|00\rangle$ and $|10\rangle$ in the second case with  $\theta_1(t_f)=0.7021\pi$ than that by the first case with $\theta_1(t_f)=0.3412\pi$. The analytically calculated phases as expected by Eq. (\ref{J0M0}) in Figs. \ref{fig6} (d)-(f)  are unchanged for small changes of detuning. The corresponding optical transition processes via higher-order Magnus terms strongly affect the phases of rotational states.  As a result, the oscillations of the analytically calculated orientations observed in Fig. \ref{fig4} (f) are suppressed by optical processes beyond the first-order Magnus description. For the excitation with extremely small detuning, i.e., $\Delta\approx0$, the exactly calculated results nicely fit the analytical predictions by satisfying the amplitude and phase conditions of the control fields.   
\subsection{Simulations with two time-delayed pulses}
\begin{figure}[!t]\centering
\resizebox{0.7\textwidth}{!}{%
 \includegraphics{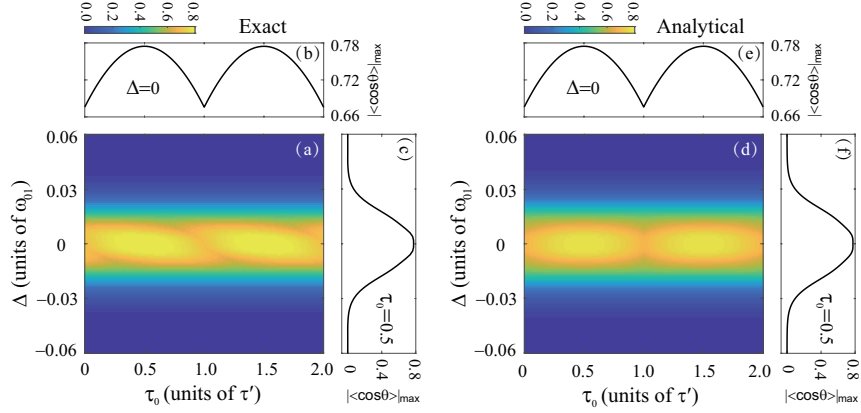}\textit{}
} \caption{Numerical simulations with two time-delayed pulses for the first condition in Eq. (\ref{ac1}) with a narrow bandwidth $\Delta\omega=0.02/\tau'$. (a) and (b) The maximum orientation values $\left|\langle\cos\theta\rangle\right|_{max}$ versus the time delay and the detuning of the pulses. The left panels (a-c) correspond to the exact simulations by using the time-dependent unitary operator in Eq. (\ref{UO}). The right panels (d-f) correspond to the analytical simulations by using Eq. (\ref{J0M0}). } \label{fig7}
\end{figure}
\begin{figure}[!t]\centering
\resizebox{0.7\textwidth}{!}{%
 \includegraphics{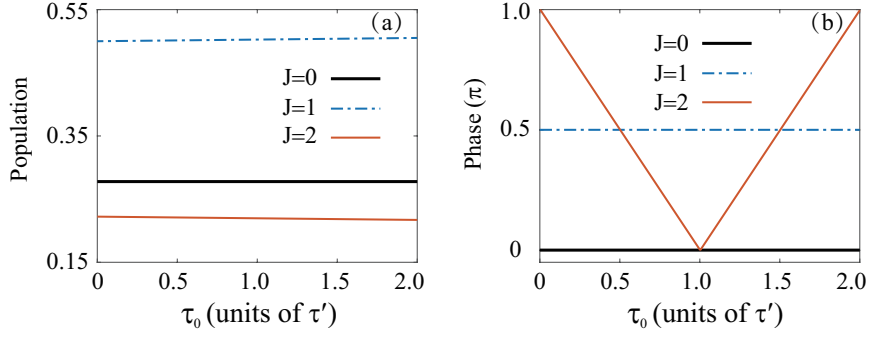}
    } \caption{The dependence of exactly calculated populations and phases, i.e., $|c_{J'}(t_f)|^2$ and $\arg[c_{J'}(t_f)]$ on the time delay $\tau_0$ for $\Delta=0$, corresponding to the orientation in Fig. \ref{fig7} (b). } \label{fig8}
\end{figure}

We further examine the scheme by considering two time-delayed terahertz pulses. To satisfy both amplitude and phase conditions, we perform the following simulations using two pulses with a narrow bandwidth of $\Delta\omega=0.02/\tau'$ with $\tau'=\pi/2\omega_{01}$.  We apply the first condition in Eq. (\ref{ac1}) to the simulations and fix the phases $\phi_1=\phi_2=0$.  Figure \ref{fig7} shows the maximum value of the orientation versus the delay time $\tau_0$ and detuning $\Delta$ of the pulses. We can see that the maximum orientation values are highly sensitive to the two laser parameters. At the excitation condition of $\Delta=0$, the orientation values can reach the theoretical maximum at $\tau_0=0.5\tau'$ and $1.5\tau'$, where the phases $\phi_1$, $\phi_2$ and $\omega_{12}\tau_0$ satisfy the phase condition in Eq. (\ref{phase}), and the exactly calculated results are in good agreements with the analytical ones.  To see how the delay time affects the orientation values, Fig. \ref{fig8} shows the populations and phases of the three rotational states versus the time delay after the pulses are turned off. We can see that the populations keep unchanged as the time delay varies, and the phase of the state $|20\rangle$ changes as expected by Eqs. (\ref{J0M0}) and (\ref{tde1}). Note that the exactly calculated results in Figs. \ref{fig7} (a)-(c) follow the analytical ones in Figs. \ref{fig7} (d)-(f) very well, only with slight differences concerning detuning. 
 \\ \indent
\begin{figure}[!t]\centering
\resizebox{0.7\textwidth}{!}{%
 \includegraphics{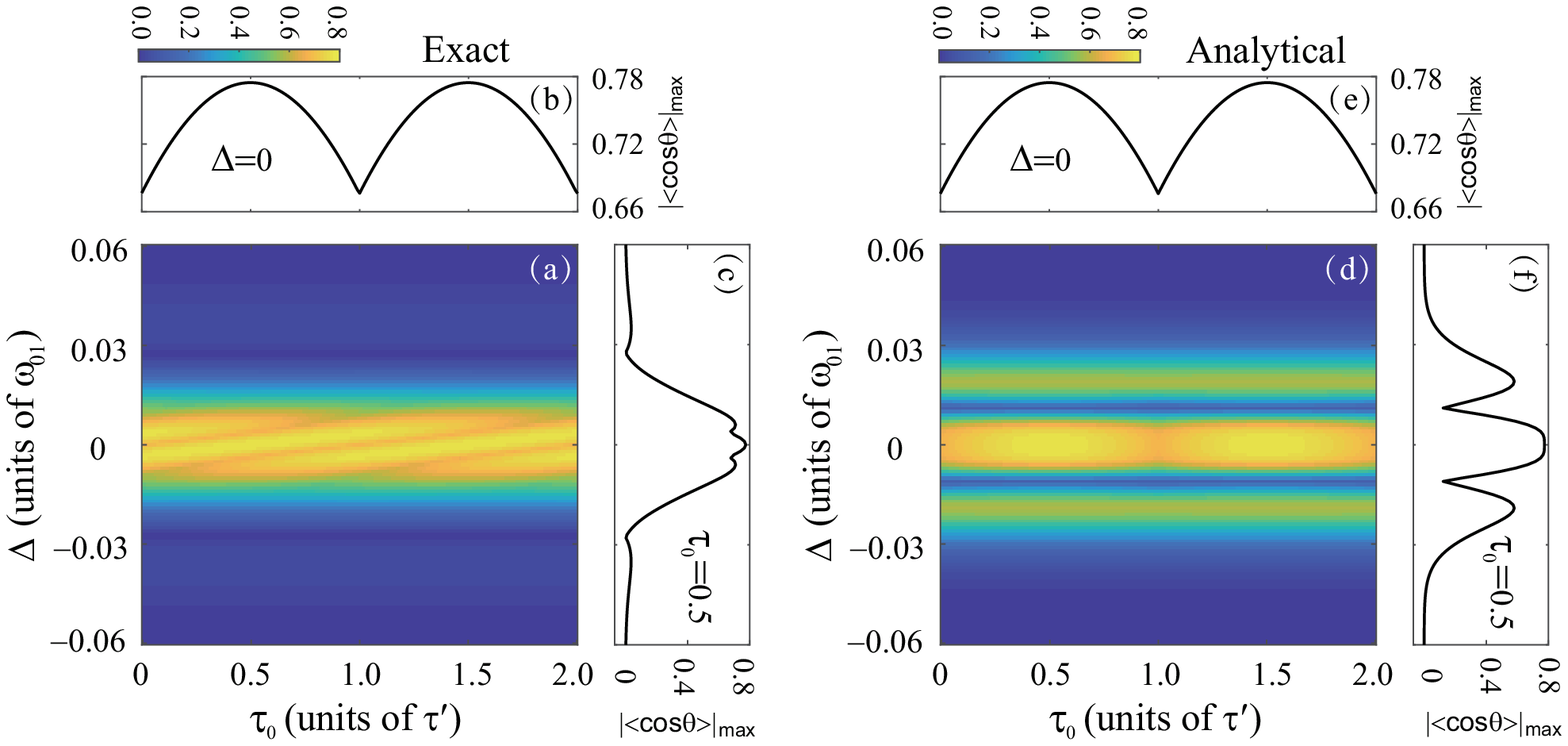}
} \caption{The same simulations as in Fig. \ref{fig7}  for the second condition in Eq. (\ref{ac2}).} \label{fig9}
\end{figure}
\begin{figure}[!t]\centering
\resizebox{0.7\textwidth}{!}{%
 \includegraphics{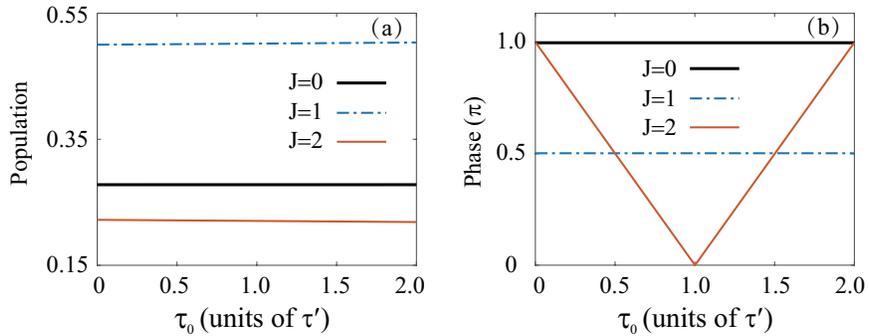}
} \caption{The same simulations as in Fig. \ref{fig8}  for the second condition in Eq. (\ref{ac2}).} \label{fig10}
\end{figure}
Figures \ref{fig9} and \ref{fig10} display the same simulations as Fig. \ref{fig7} and \ref{fig8} by satisfying the second condition in Eq. (\ref{ac2}). In both simulations, the maximum values of orientation exhibit strong dependences on the laser parameters,  i.e., detuning and time delay. For the resonant excitation with $\Delta=0$, the exactly calculated orientation in Fig. \ref{fig9} (b) is consistent with the analytical result in Fig. \ref{fig9} (e). The corresponding populations and phases in Fig. \ref{fig10} show the same behaviors as that in Fig. \ref{fig8}, i.e., changing the delay time leads to a phase change of the state $|20\rangle$.  When we look at the orientation and its dependence on the detuning for a given time delay, the exact results are different from the analytical ones. The latter shows clear oscillations concerning the detuning in Fig. \ref{fig9} (f), whereas this phenomenon is strongly suppressed in Fig. \ref{fig9} (c). As analyzed for the second case of the zero-delay in Fig. \ref{fig6}, the underlying physics can be attributed to the strong laser coupling between the states $|00\rangle$ and $|20\rangle$, where the high-order Magnus terms play roles for both populations and phases. As a result, the second amplitude and phase conditions of Eqs. (\ref{ac2}) and (\ref{phase}) can only be used for extremely small detuning. \\ \indent 
Note that in the above simulations, two kinds of conditions in Eqs. (\ref{ac1})-(\ref{phase}) are examined for the case of $j=0$, i.e., $\theta_1(t_f)$ and $\theta_2(t_f)$ within the range of $[0, \pi]$. In principle, the theoretical conditions of Eqs. (\ref{ac1})-(\ref{phase}) can be analyzed for the cases of $j>0$, for which the Rabi couplings  between states (i.e., the off-diagonal elements in Eq. (\ref{HI})) become stronger than that for $j=0$. As a result, the effect of optical excitations by high-order Magnus terms on the orientation may become more pronounced. Therefore the requirements that lead to the maximum degree of orientation are strictly limited for using the two conditions.  To that end,  the experimental realization would be more accessible when $j=0$ in Eqs. (\ref{ac1})-(\ref{phase}). 
 \section{Conclusion} \label{conl}
 We performed a combined analytical and numerical analysis to find optimal control fields for achieving the maximum field-free molecular orientation within a three-state model. Using the first-order Magnus expansion to the time-dependent unitary operator, we obtained an analytical solution for a three-state-time-dependent wave function.  We then derived amplitude and phase conditions, i.e.,  Eqs. (\ref{ac1}) -(\ref{phase}) for the control fields, resulting in the theoretical maximum orientation value with an optimal combination of populations and phases for the three rotational states. Based on the amplitude and phase conditions, we suggested a quantum coherent control scheme and successfully applied it to the linear polar molecules HCN with four different simulations. Multiple optimal solutions exist and we investigated the frequency-domain landscape of the optimal terahertz pulses. As a result, we showed how pure rotational ladder-climbing excitation could generate an optimal three-state molecular orientation by suppressing simultaneous multiphoton excitation processes. This work provides an essential reference to the optimization of the populations and phases of rotational states, leading to the maximum degree of orientation within a three-state model. Our analysis that was based on the initial rotational state $|00\rangle$, can be extended to molecules initially in other rotational states. Then, the optimal amplitude and phase conditions can be derived for generating the maximum orientation of molecules at finite temperatures.  

\begin{acknowledgements}
 This work was supported by the National Natural Science Foundations of China (NSFC) under Grant No. 61973317.
\end{acknowledgements}

\end{document}